\documentclass[twocolumn,showpacs,
superscriptaddress,preprintnumbers,amsmath,amssymb]{revtex4}

\usepackage{graphicx}% Include figure files
\usepackage{dcolumn}% Align table columns on decimal point
\usepackage{bm}% bold math

\def\Vec#1{\mbox{\boldmath$#1$}}

\begin{document}
  
  \preprint{}
  
  \title{
    Relaxation Time and Relaxation Function of Quark-Gluon Plasma with Lattice QCD
  }
  
  \author{Kyosuke Tsumura}
  \affiliation{
    Analysis Technology Center,
    Fujifilm Corporation,
    Kanagawa 250-0193, Japan
  }
  
  \begin{abstract}
    We propose a method
    which enables a
    QCD-based calculation
    of a relaxation time
    for a dissipative current
    in the causal and dissipative hydrodynamic equation
    derived by Israel and Stewart.
    We point out that
    the Israel-Stewart equation is not unique
    as a causal and dissipative hydrodynamic equation,
    and the form of the causal and dissipative hydrodynamic equation
    is determined by the shape of a spectral function reflecting
    the properties of elementary excitations in the system we consider.
    
    Our method utilizes a relaxation function,
    which can be calculated from QCD
    using the linear response theory.
    We show that the relaxation function
    can be derived from
    a spectral function for a microscopic representation of the dissipative current.
    We also show that the Israel-Stewart equation is acceptable
    only as long as the calculated relaxation function is
    approximated well by a exponentially damping function,
    and the relaxation time
    can be obtained as its damping time constant.
    Taking a baryon-number dissipative current
    of a plasma consisting of charm quarks and gluons
    as a simple example,
    we present the first calculation of the relaxation function
    with use of the spectral function derived
    employing the quenched lattice QCD together with the maximum
    entropy method.
    The calculated relaxation function
    shows a strongly-oscillation damping behaviour
    due to
    the charmed vector hadron
    $J/\Psi$ surviving
    above the deconfinement phase transition temperature in QCD.
    This result suggests that
    the applicability of the Israel-Stewart equation
    to the baryon-number dissipative current
    of the charm quark-gluon plasma
    is quite doubtful.
    We present an idea for
    the improvement of the Israel-Stewart equation
    by deriving the hydrodynamic equation
    consistent with the strongly-oscillation damping relaxation function.
  \end{abstract}

  \date{\today}
  \maketitle
  
  \section{Introduction\label{sec:001}}
  The time evolution of the hot
  matter of quarks and gluons
  created in the Relativistic Heavy Ion Collider (RHIC)
  experiments at Brookhaven National Laboratory
  is well described by the relativistic hydrodynamic equation with
  \textit{almost no} dissipative effects \cite{qgp001,qgp002,qgp003}.
  This fact means that
  the quark-gluon plasma (QGP)
  slightly above the deconfinement phase transition temperature $T_C$
  behaves as
  an almost ideal
  fluid,
  of which viscosity and heat conductivity, i.e., transport coefficients, are
  negligiblely
  small.
  This interesting non-equilibrium property of QGP
  as an almost ideal fluid
  is considered
  to be one of the properties of a strongly interacting system,
  and
  consistent with
  hadronic excitations surviving in QGP
  shown by several groups using the lattice QCD \cite{mem003,mem004,mem005,mem006}.
  
  In order to reach a full understanding
  of this non-equilibrium property of QGP,
  we need to study the RHIC experimental data
  using the relativistic hydrodynamic equation
  with
  \textit{finite} dissipative effects.
  
  At present,
  a good candidate
  for the relativistic hydrodynamic equation with dissipative effects
  is considered to be
  the Israel-Steart equation \cite{2nd001,2nd002}
  instead of
  the standard equations
  by Eckart \cite{1st001} and Landau and Lifshitz \cite{1st002}.
  This is because
  the Israel-Stewart equation holds the causality
  in contrast with
  the Eckart and Landau-Lifshitz equations.
  
  To assure the causality,
  the Israel-Stewart equation
  contains higher derivative terms
  with relaxation times, i.e.,
  new parameters other than transport coefficients,
  which are dependent on materials we consider.
  An important matter is
  to determine these relaxation times
  from microscopic dynamics governing the materials.
  
  In 1979, Israel and Stewart \cite{2nd002} calculated
  the relaxation times of a rarefied gas
  by
  using the relativistic Boltzmann equation
  with the Grad's fourteen moment approximation \cite{grad}.
  These quantities have been used by several groups \cite{dis001,dis002,dis003}
  to carry out
  (2+1)-dimensional hydrodynamic simulations
  based on the Israel-Stewart equation.
  
  In 2008, Natsuume and Okamura \cite{ads001} derived the relaxation times
  of a strongly interacting supersymmetric gauge-theory plasma
  by utilizing the dispersion relations of the hydrodynamic modes
  obtained with use of the AdS/CFT duality.
  However,
  a calculation of the relaxation times of QGP
  based on QCD
  is unfortunately not yet done.
  
  The aim of this paper is
  (1) to construct a method
  which enables us to determine the relaxation times
  from the microscopic theory including QCD,
  and (2) to present the first lattice QCD calculation
  of the relaxation times of QGP
  on the basis of this method.
  
  To construct the method,
  we utilize a \textit{relaxation function},
  which
  gives a linear but non-local (non-Markovian) relation between
  an external field applied to the system
  and
  the current induced in the system \cite{lrt001,lrt002}.
  The relaxation functions
  contain full information
  about non-equilibrium dissipative processes,
  e.g., transport coefficients and relaxation times.
  Moreover a microscopic calculation
  of the relaxation functions
  is straightforward using the linear response theory.
  
  In this paper,
  we first express the relaxation function in terms of
  a \textit{spectral function}
  for a conserved current,
  which can be calculated from the microscopic dynamics
  by a quantum-field theoretic method.
  This expression tells us that
  the relaxation function is determined
  by a whole behaviour of the spectral function
  in contrast to the transport coefficient
  defined as the low-frequency and long-wavelength limit
  of the spectral function.
  We show that,
  only
  as long as
  the calculated relaxation function
  shows an exponentially damping behaviour,
  the damping time constant
  can be identified
  as
  the relaxation time in the Israel-Stewart equation.
  Taking a baryon-number dissipative current
  of a plasma consisting of charm quarks and gluons
  (referred to as the charm quark-gluon plasma later on)
  as a simple example,
  we calculate the relaxation function from the spectral function obtained
  using the lattice QCD together with the maximum entropy method (MEM)
  \cite{mem001}.
  The obtained relaxation function shows
  a strongly-oscillation damping behaviour
  instead of an exponentially damping
  one,
  due to the charmed vector hadron
  $J/\Psi$ surviving above $T_C$.
  This disagreement suggests that
  the applicability of the Israel-Stewart equation
  to the baryon-number dissipative current of the charm quark-gluon plasma
  is doubtful.
  We show
  an idea for the improvement
  of the Israel-Stewart equation
  by deriving the hydrodynamic equation
  consistent with the strongly-oscillation damping relaxation function.
  
  The conclusion of this paper is that
  the Israel-Stewart equation
  is not unique as the causal and dissipative hydrodynamic equation,
  and the form of the causal and dissipative hydrodynamic equation
  applicable to QGP
  is determined by the shape of the spectral function
  reflecting the properties of
  hadronic excitations in QGP.
  This indicates
  an important link between
  the non-equilibrium properties of QGP
  and the hadron spectroscopy at finite temperature.
  
  This paper is organized as follows.
  In section \ref{sec:002},
  we construct a method
  which enables a microscopic calculation of the relaxation times
  in the Israel-Stewart equation.
  For this purpose,
  the relaxation functions are utilized.
  In section \ref{sec:003},
  we present the first calculation of the relaxation function
  for the baryon-number dissipative current in QGP,
  with use of the Monte Carlo simulation based on the lattice QCD.
  We discuss
  whether the exponentially damping relaxation function
  of the Israel-Stewart equation can be justified
  from a microscopic point of view.
  Section \ref{sec:004} is devoted to the summary and the concluding remarks.
  In appendix \ref{sec:005},
  taking a shear viscous dissipative process as a typical example,
  we show the detailed derivation of the microscopic representation
  of the relaxation function.

  \section{Formula for Relaxation Function\label{sec:002}}
  In this section,
  we give a method
  which enables us to determine the relaxation time
  from the microscopic theory including QCD.
  Our discussion
  is based on
  the relaxation function
  established in the phenomenological relaxation theory \cite{lrt001}.
  The microscopic representation of the relaxation function
  is given in the linear response theory \cite{lrt001,lrt002}.
  
  \subsection{Non-Markovian Constituent Equation
    Derived from Israel-Stewart Equation\label{sec:002A}}
  To explain the physical meaning of the relaxation functions,
  let us start from the Israel-Stewart equation for a simple system,
  where the conserved quantities are the energy and the momentum,
  and the dissipative effect comes solely from the shear viscosity.
  We denote the energy-momentum tensor by
  \begin{eqnarray}
    \label{eq:2-001}
    T^{\mu\nu} = \epsilon \, u^\mu \, u^\nu - p \, \Delta^{\mu\nu} + \pi^{\mu\nu},
  \end{eqnarray}
  where
  $\epsilon = \epsilon(T)$ and $p = p(T)$ are
  the internal energy and the pressure at the temperature $T$,
  respectively,
  $u^\mu$ the flow velocity,
  $\pi^{\mu\nu}$ the shear viscous stress
  and
  $\Delta^{\mu\nu} \equiv g^{\mu\nu} - u^\mu \, u^\nu$
  the projection operator
  with $g^{\mu\nu} = \mathrm{diag}(+1,-1,-1,-1)$.
  Here $u^\mu$ is normalized as $u_\mu \, u^\mu = 1$,
  and $\pi^{\mu\nu}$ is satisfied with
  $\pi^{\mu\nu} = \pi^{\nu\mu}$ and
  $u_\mu \, \pi^{\mu\nu} = \pi^\mu_{\,\,\,\mu} = 0$,
  which are reproduced by
  $\pi^{\mu\nu} = \Delta^{\mu\nu\rho\sigma} \,
  \tilde{\pi}_{\rho\sigma}$
  with the symmetric traceless projection operator
  $\Delta^{\mu\nu\rho\sigma} \equiv 1/2 \, (
  \Delta^{\mu\rho} \, \Delta^{\nu\sigma} + \Delta^{\mu\sigma} \, \Delta^{\nu\rho} - 2/3 \,
  \Delta^{\mu\nu} \, \Delta^{\rho\sigma})$
  and the shear viscous stress before the projection
  $\tilde{\pi}_{\rho\sigma}$.
  It is noted that the Israel-Stewart equation
  is the system of the nine differential equations
  with respect to the nine independent variables, $T$, $u^\mu$ and
  $\pi^{\mu\nu}$.
  The equations consist of
  the continuity equation,
  \begin{eqnarray}
    \label{eq:2-002}
    \partial_\mu T^{\mu\nu} = 0,
  \end{eqnarray}
  and the relaxation equation \cite{2nd001,2nd002},
  \begin{eqnarray}
    \label{eq:001}
    \tau_\pi \, D \tilde{\pi}_{\rho\sigma} + \tilde{\pi}_{\rho\sigma} = 2 \,
    \eta \, \partial_\rho u_\sigma,
  \end{eqnarray}
  where $\eta$ and $\tau_\pi$ denote
  the shear viscosity and the relaxation time of
  the shear viscous dissipative process, respectively,
  and $D \equiv u^\mu\, \partial_\mu \equiv \partial/\partial\tau$ with the
  proper time $\tau$.
  
  The solution of Eq.(\ref{eq:001}) for the shear viscous stress reads
  \begin{eqnarray}
    \label{eq:002}
    \tilde{\pi}_{\rho\sigma}(\tau)
    = \int\!\!\mathrm{d}\tau^\prime \, \Bigg[ \theta(\tau -
    \tau^\prime)
    \, \frac{\eta}{\tau_\pi} \,
    \mathrm{e}^{- (\tau-\tau^\prime)/\tau_\pi} \Bigg] \,
    2 \, \partial_\rho u_\sigma(\tau^\prime).
  \end{eqnarray}
  Equation (\ref{eq:002}) tells us that
  $\pi^{\mu\nu} = \Delta^{\mu\nu\rho\sigma} \, \tilde{\pi}_{\rho\sigma}$
  at the proper time $\tau$
  depends on the history of the external field $2 \, \partial_\rho u_\sigma$
  from past to present,
  and
  the weighting function
  is given by
  $\frac{\eta}{\tau_\pi} \,
  \mathrm{e}^{- (\tau-\tau^\prime)/\tau_\pi}$.
  Notice that
  Eq.(\ref{eq:002}) is equivalent to that derived
  in the Stewart's non-local thermodynamics \cite{non001},
  and has been recently rediscovered by Koide \textit{et al.} \cite{non002}.
  The meaning of Eq.(\ref{eq:002}) becomes clearer
  from the viewpoint of
  the phenomenological relaxation theory \cite{lrt001},
  where the linear but non-Markovian constituent equation,
  \begin{eqnarray}
    \tilde{\pi}_{\rho\sigma}(\tau) =
    \int\!\!\mathrm{d}\tau^\prime \, R(\tau-\tau^\prime) \, 2 \,
    \partial_\rho u_\sigma(\tau^\prime),
  \end{eqnarray}
  is well established
  and $R(\tau)$ is identically the relaxation function.
  Therefore, we find that
  the Israel-Stewart causal hydrodynamics is
  equivalent to the special case of the phenomenological relaxation theory
  with
  \begin{eqnarray}
    \label{eq:2-003}
    R(\tau) = \theta(\tau) \, \frac{\eta}{\tau_\pi} \,
    \mathrm{e}^{- \tau/\tau_\pi},
  \end{eqnarray}
  i.e., the exponentially damping relaxation function.
  
  \subsection{Microscopic Representation of Relaxation Function
    Based on Spectral Function\label{sec:002B}}
  We first notice that
  the linear response theory \cite{lrt001,lrt002}
  enables us to calculate the relaxation function
  directly from the microscopic dynamics.
  We now consider
  the relaxation function of
  a generic dissipative process,
  \begin{eqnarray}
    \label{non-Markov}
    J(t ,\, \Vec{x}) = \int\!\!\mathrm{d}u \,
    \mathrm{d}^3\Vec{y} \,
    R(t-u ,\, \Vec{x}-\Vec{y}) \,
    X(u ,\, \Vec{y}),
  \end{eqnarray}
  where $X(t ,\, \Vec{x})$ denotes the external field,
  $J(t ,\, \Vec{x})$ the current induced by $X(t ,\, \Vec{x})$,
  and $R(t ,\, \Vec{x})$ the relaxation function.
  The linear response theory
  allows that
  the microscopic derivation of the relaxation function
  is straightforward.
  In fact, as shown in appendix \ref{sec:005},
  we can derive the following relation,
  \begin{eqnarray}
    \label{eq:003}
    R(t ,\, \Vec{x})
    = \theta(t) \,
    \int_{-\infty}^0\!\!\mathrm{d}s \,
    \theta(t-s) \,
    i \,
    \langle \, [ \, \hat{J}(t ,\, \Vec{x}) \,,\,
      \hat{J}(s ,\, \Vec{0}) \, ] \,
    \rangle,
  \end{eqnarray}
  where
  $\hat{J}(t ,\, \Vec{x})$ is
  a Heisenberg operator and gives
  the microscopic representation for
  the induced macroscopic current $J(t ,\, \Vec{x})$,
  and the bracket indicates the thermal average.
  Notice that the integrand of the right hand side of Eq.(\ref{eq:003})
  is the retarded Green function,
  which has the following spectral representation,
  \begin{eqnarray}
    \label{eq:004}
    \theta(t) \, i \, \langle \, [ \, \hat{J}(t ,\, \Vec{x}) \,,\,
      \hat{J}(0 ,\, \Vec{0}) \, ] \,
    \rangle \nonumber\\
    {} = \int\!\!\frac{\mathrm{d}^4k}{(2\pi)^4} \,
    \mathrm{e}^{-ik^0 t+i
      \mbox{\boldmath$\scriptstyle k$}\cdot
      \mbox{\boldmath$\scriptstyle x$}
    } \,
    \int\!\!\mathrm{d}\omega \, \frac{A(\omega,\,\Vec{k})}{\omega - k^0 - i \, \varepsilon},
  \end{eqnarray}
  where $A(\omega,\,\Vec{k})$ is the spectral function
  and
  $\varepsilon$ the infinitesimal positive constant.
  Equations (\ref{eq:003}) and (\ref{eq:004}) lead us to
  the relation between $R(t ,\, \Vec{x})$ and $A(\omega ,\, \Vec{k})$
  given by
  \begin{eqnarray}
    \label{eq:005}
    R(t ,\, \Vec{x}) = 2 \, \theta(t) \,
    \int\!\!\frac{\mathrm{d}\omega\mathrm{d}^3\Vec{k}}{(2\pi)^4} \,
    \mathrm{e}^{-i\omega t+i
      \mbox{\boldmath$\scriptstyle k$}\cdot
      \mbox{\boldmath$\scriptstyle x$}
    } \,
    \frac{\pi}{\omega} \, A(\omega,\,\Vec{k}).
  \end{eqnarray}
  This formula assures that the relaxation function can be obtained
  as the Fourier transformation of $\pi \, A(\omega,\,\Vec{k}) / \omega$,
  i.e., the real part of
  the complex admittance \cite{lrt001},
  which is defined by
  \begin{eqnarray}
    \label{eq:admittance}
    \tilde{R}(k^0,\,\Vec{k}) &\equiv&
    \int\!\!\mathrm{d}t\,\mathrm{d}^3\Vec{x}\,R(t,\,\Vec{x})\,
    \mathrm{e}^{ik^0t-i
      \mbox{\boldmath$\scriptstyle k$}\cdot
      \mbox{\boldmath$\scriptstyle x$}
    }\nonumber\\
    &=& \int\!\!\mathrm{d}\omega \,
    \frac{-i}{\omega - k^0 - i \, \varepsilon} \,
    \frac{A(\omega,\,\Vec{k})}{\omega}.
  \end{eqnarray}
  Here we
  notice
  the two important properties
  concerning the complex admittance $\tilde{R}(k^0,\,\Vec{k})$:
  (i) The low-frequency and long-wavelength limit
  of the complex admittance,
  $\tilde{R}(0,\,\Vec{0})
  = \pi \, A(\omega,\,\Vec{0}) / \omega \big|_{\omega=0}$,
  gives the transport coefficients.
  In fact, with use of this representation,
  many groups \cite{tra001,tra002,tra003,tra004}
  have calculated the transport coefficients of QGP
  employing the lattice QCD.
  (ii) Using $\tilde{R}(k^0,\,\Vec{k})$,
  we can write down the generic relaxation equation
  equivalent to the non-Markovian constituent equation (\ref{non-Markov}),
  \begin{eqnarray}
    \label{eq:2-004}
    \tilde{R}^{-1}(k^0= +i\,\partial_t,\,\Vec{k}=-i\,\Vec{\nabla})
    \, J(t,\, \Vec{x}) = X(t,\, \Vec{x}).
  \end{eqnarray}
  By combining Eqs.(\ref{eq:admittance}) and (\ref{eq:2-004}),
  we can construct the relaxation equation
  from a given spectral function.
  
  The above observation tells us the following three points:
  (a) The relaxation function in QGP can be calculated from the
  spectral function
  for the microscopic current of QCD.
  (b) The Israel-Stewart equation
  can be valid
  for the description of QGP
  only as long as the relaxation function calculated
  in the microscopic way
  is an exponentially damping function,
  and the relaxation time of QGP
  can be obtained as
  its damping time constant.
  (c) If the relaxation function
  calculated from QCD
  is different from that derived from the Israel-Stewart equation,
  we should carry out the causal hydrodynamic calculation
  by incorporating the relaxation equation (\ref{eq:2-004}) or
  the non-Markovian constituent equation (\ref{non-Markov})
  into the continuity equation.
  
  \subsection{Spectral Function Consistent with Israel-Stewart Equation\label{sec:002C}}
  It is interesting to investigate
  which spectral functions reproduce the exponentially damping
  relaxation function in the Israel-Stewart equation.
  Taking the shear viscous dissipative process
  at the rest frame
  as a typical example,
  we find it to be a Lorentzian function,
  \begin{eqnarray}
    \label{eq:2-005}
    \frac{\pi}{\omega} \, A(\omega ,\, \Vec{k})
    = \frac{\eta}{1 + \omega^2 \, \tau^2_\pi}.
  \end{eqnarray}
  In fact,
  it can be checked that
  Eqs.(\ref{eq:005}) and (\ref{eq:2-005}) give
  the exponentially damping relaxation function,
  \begin{eqnarray}
    \label{eq:2-006}
    R(t,\,\Vec{x}) = \theta(t) \, \frac{\eta}{\tau_\pi} \,
    \mathrm{e}^{- t/\tau_\pi} \, \delta^3(\Vec{x}),
  \end{eqnarray}
  and the correspondent complex admittance reads
  \begin{eqnarray}
    \label{eq:2-007}
    \tilde{R}(k^0,\,\Vec{k}) = \frac{\eta}{1 - i \, k^0 \,\tau_\pi }.
  \end{eqnarray}
  It is noteworthy that Eq.(\ref{eq:2-004}) with this complex admittance reproduces
  the Israel-Stewart-type relaxation equation,
  \begin{eqnarray}
    \label{eq:2-008}
    (1 + \tau_\pi \, \partial_t) J(t,\,\Vec{x}) = \eta \, X(t,\,\Vec{x}).
  \end{eqnarray}
  Here we replace $\partial_t$ with $D = u^\mu \partial_\mu = \partial/\partial\tau$
  in order to
  obtain the equation at an arbitrary
  frame
  from that at the rest frame.
  Therefore it is found that
  the ansatz of
  the Lorentzian-type spectral function shown in Eq.(\ref{eq:2-005})
  is an essential point of the Israel-Stewart equation.
  Notice that
  there is no reason to believe that
  the spectral function for the current of QCD is approximated well by
  the Lorentzian function
  in the low-frequency and long-wavelength region.

  \section{Results from Lattice QCD\label{sec:003}}
  In this section,
  we demonstrate the first lattice QCD calculation
  of the relaxation function.
  From the viewpoint of availability of lattice data,
  we choose the relaxation function
  for the baryon-number dissipative process.
  The continuity equation reads
  \begin{eqnarray}
    \label{eq:3-001}
    \partial_\mu N^\mu \equiv \partial_\mu (n \, u^\mu + \nu^\mu) = 0,
  \end{eqnarray}
  and the non-Markovian constituent equation
  \begin{eqnarray}
    \label{eq:3-002}
    \nu^\mu(t ,\, \Vec{x})
    = \Delta^{\mu\nu}(t ,\, \Vec{x}) \, \int\!\!\mathrm{d}u \,
    \mathrm{d}^3\Vec{y} \,
    R(t-u ,\, \Vec{x}-\Vec{y})\nonumber\\
    \times \partial_\nu (\mu_\mathrm{B}/T)(u ,\, \Vec{y}),
  \end{eqnarray}
  where $n$ and $\mu_\mathrm{B}$ denote the baryon-number density
  and the baryon-number chemical potential, respectively.
  This relaxation function can be calculated from
  the spectral function of the spatial components of the
  baryon-number current,
  $J^\mu = \sum_{f=1}^{N_\mathrm{F}} \, \sum_{c=1}^{N_\mathrm{C}} \,
  \bar{\psi}_{fc} \, \gamma^\mu \, \psi_{fc}$,
  where
  $\psi_{fc}$ denotes the $f$-flavour $c$-colour quark field,
  $\gamma^\mu(\mu=0,1,2,3)$ the gamma matrix,
  $\bar{\psi}_{fc} = \psi^\dagger_{fc} \gamma^0$,
  $N_\mathrm{F}$ the number of the flavour
  and $N_\mathrm{C}(=3)$ the number of the colour, respectively.
  As a simple example,
  we consider the single-flavour deconfined system
  of which constituents are gluons and charm quarks ($N_\mathrm{F}=1$),
  i.e., the charm quark-gluon plasma.
  Furthermore we treat
  a spatially integrated relaxation function,
  \begin{eqnarray}
    \label{eq:3-003}
    R(t) \equiv \int\!\!\mathrm{d}^3\Vec{x} \, R(t,\,\Vec{x})
    = 2 \, \theta(t) \, \int\!\!\frac{\mathrm{d}\omega}{2\pi} \,
    \mathrm{e}^{-i\omega t} \,
    \frac{\pi}{\omega} \, A(\omega),
  \end{eqnarray}
  where
  $A(\omega) \equiv A(\omega,\,\Vec{0})$ is
  a zero-momentum projected spectral function.
  This relation indicates that
  the calculation of $R(t)$ requires
  an explicit form of $A(\omega)$.
  Therefore we need to reconstruct
  $A(\omega)$ from the temporal correlation function,
  \begin{eqnarray}
    \label{eq:3-004}
    D(\tau) = 1/3 \, \sum_{i=1}^3 \,
    \int\!\!\mathrm{d}^3\Vec{x} \, \langle \, J^i(-i\tau ,\, \Vec{x}) \, J^i(0,\,\Vec{0})
    \, \rangle,
  \end{eqnarray}
  obtained using Monte Carlo simulation based on the lattice QCD.
  To get the full information about $A(\omega)$,
  we utilize MEM \cite{mem001}.
  
  \subsection{Outline of MEM Analysis\label{sec:003A}}
  Here we shall give a brief account of MEM.
  The temporal correlation function $D(\tau)$ and the
  associated $A(\omega)$ are related by the Laplace transformation,
  \begin{eqnarray}
    \label{eq:3-005}
    D(\tau) = \int_0^\infty\!\!\mathrm{d}\omega \,
    K(\tau,\,\omega) \, A(\omega),
  \end{eqnarray}
  where
  $K(\tau,\,\omega) \equiv \cosh[\omega \, (\tau - 1/2T)] /
  \sinh(\omega/2T)$ is a kernel function.
  Since the number of data points in the temporal direction is finite,
  the direct inverse Laplace transformation from
  $D(\tau)$ to $A(\omega)$ is an ill-posed problem.
  MEM is a method to
  overcome
  this difficulty on the basis of the Bayes' theorem
  in the theory of statistical inference.
  Using MEM, we can infer $A(\omega)$
  as the most probable one
  which
  achieves a balance between
  reproducing the lattice data $D(\tau)$
  and
  being kept not so far from the default model $m(\omega)$, defined by
  a plausible form of $A(\omega)$.
  MEM has proved to be a powerful tool to investigate $A(\omega)$ of mesons and baryons
  at both of $T = 0$ \cite{mem001,mem002} and $T \neq 0$
  \cite{mem003,mem004,mem005,mem006}.
  
  Recently, the author \cite{mem007} has improved MEM to avoid a numerical
  instability present in
  the study of the shear viscosity of the gluon plasma.
  An essential point of the improved MEM analysis
  is that
  the Laplace transformation combining
  $D(\tau)$ and $A(\omega)$ is converted as follows,
  \begin{eqnarray}
    \label{eq:006}
    D(\tau) = \int_0^\infty\!\!\mathrm{d}\omega \, \Bigg[ K(\tau,\,\omega) \,
      \frac{\omega}{\pi} \Bigg] \, \Bigg[ \frac{\pi}{\omega} \,
    A(\omega) \Bigg].
  \end{eqnarray}
  Regarding
  $\omega \, K(\tau,\,\omega) / \pi$
  as a new
  kernel
  function,
  we can infer the real part of the complex admittance
  $\pi \, A(\omega)/\omega$ directly from $D(\tau)$.
  Further,
  in the study of the scalar tetraquark \cite{nex001},
  the extension of this method, such as
  $D(\tau) = \int_0^\infty\!\!\mathrm{d}\omega \,
  [K(\tau,\,\omega) \Lambda(\omega)] \, [A(\omega)/\Lambda(\omega)]$
  with $\Lambda(\omega) \equiv \omega^8$,
  has
  also been
  used to reduce the uncertainty of the inferred $A(\omega)$.
  Aarts \textit{et al.} \cite{tra004} independently have
  prepared
  this method,
  and have applied this method
  to calculate the electrical conductivity of QGP.
  
\begin{figure}[tHb]
    \begin{center}
      \begin{minipage}{1.0\linewidth}
        \includegraphics[width=\linewidth]{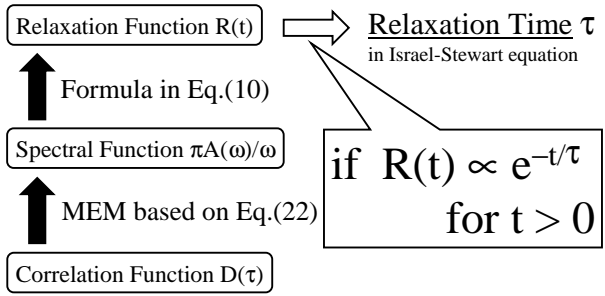}
      \end{minipage}
    \end{center}
    \caption{
      The schematic overview to calculate the relaxation times
      in the Israel-Stewart causal and dissipative hydrodynamic equation
      employing the lattice QCD together with the maximum entropy
      method (MEM).
      The correlation function can be calculated in lattice.
    }
    \label{fig:000}
  \end{figure}

  In Fig.\ref{fig:000},
  we show the schematic overview
  which summarizes the way to
  obtain the relaxation function
  and the relaxation time of QGP
  on the basis of the lattice QCD
  and the improved MEM analysis.
  
  Then we explain the default model used in the present analysis.
  The default model $m(\omega)$ is usually given
  as the form predicted
  by the perturbative
  QCD calculation valid in the high-energy region, $\omega
  \rightarrow \infty$.
  Due to the tree-level result,
  $A(\omega) = m_0 \, \omega^2$ with $m_0 \equiv 1/4\pi^2$,
  the default model reads
  \begin{eqnarray}
    \label{eq:3-006}
    m(\omega) = m_\mathrm{mono}(\omega) \equiv \pi \, m_0 \, \omega.
  \end{eqnarray}
  Such a monotonically increasing default model
  seems suitable
  to avoid the artificial peaks
  due to the choice of the default model,
  and
  using $m_\mathrm{mono}(\omega)$
  we can identify
  any peak structure of the inferred $A(\omega)$
  as that derived from the lattice data not the default model.
  In this paper, we are interested in
  the applicability of the Israel-Stewart equation to QGP,
  i.e., the validity of the ansatz that
  the spectral function for the conserved currents
  is approximated by a Lorentzian function shown in Eq.(\ref{eq:2-005}).
  To test this ansatz,
  we carry out the MEM analysis based on not only
  the monotonically increasing default model
  $m_\mathrm{mono}(\omega)$
  but also the following Lorentzian-type one especially in the low-energy region,
  \begin{eqnarray}
    \label{eq:007}
    m(\omega) = m_\mathrm{lore}(\omega) \equiv
    \frac{1}{2} \, \Big(1 - \tanh \frac{\omega -
      \omega_0}{\Delta\omega}\Big) \, \frac{\chi}{1 + \omega^2 \,
      \tau_J^2}\nonumber\\
    {} + \frac{1}{2} \, \Big(1 + \tanh \frac{\omega -
      \omega_0}{\Delta\omega}\Big) \, \pi \, m_0 \, \omega,
  \end{eqnarray}
  where $\omega_0$ and $\Delta \omega$ denote
  the center value and width of the border
  between the strongly interacting low-energy region
  and the asymptotically free high-energy one,
  respectively,
  and
  $\chi$ and $\tau_J$
  denote
  the transport coefficient and relaxation time
  of the baryon-number dissipative process, respectively.
  If $\pi \, A(\omega) / \omega$ inferred
  by the MEM analysis based on $m(\omega) = m_\mathrm{lore}(\omega)$
  is quite different from
  the Lorentzian function in the low-frequency region,
  the applicability of the Israel-Stewart equation to QGP is doubtful.
  
  \subsection{Numerical Result and Discussion\label{sec:003B}}
  The system considered here
  is the charm quark-gluon plasma
  at $T > T_C$.
  We use the quenched approximation in calculating $D(\tau)$.
  A detailed set up of the lattice QCD numerical simulation
  is identical to that in Ref.\cite{mem006}.
  We use the naive plaquette gauge action
  and the standard Wilson quark action
  with the gauge coupling constant $\beta = 7.0$,
  the bare anisotropy $\zeta_0 = 3.5$
  and
  the fermion anisotropy $\gamma_F = \kappa_\tau / \kappa_\sigma = 3.476$
  with the spatial (temporal) hopping parameter
  $\kappa_\sigma = 0.08285$ ($\kappa_\tau$).
  These parameters reproduce
  the anisotropic lattice
  $a_\sigma / a_\tau = 4$
  with the temporal (spatial) lattice spacing
  $a_\tau = 9.75 \times 10^{-3} \, \mathrm{fm}$ ($a_\sigma$)
  in the quenched-level simulation \cite{mem003}.
  The adopted lattice size is $20^3 \times 46$,
  where the spatial size $L_\sigma = 0.78 \, \mathrm{fm}$
  and the temperature $T = 1.62 \, T_C$.
  This spatial size seems small.
  By varying the spatial boundary condition,
  however,
  we have checked that $L_\sigma$ is large enough to
  simulate the charm quark-gluon plasma in lattice \cite{mem006}.
  The number of the gauge configurations is 100.
  
\begin{figure}[tHb]
    \begin{center}
      \begin{minipage}{1.0\linewidth}
        \includegraphics[width=\linewidth]{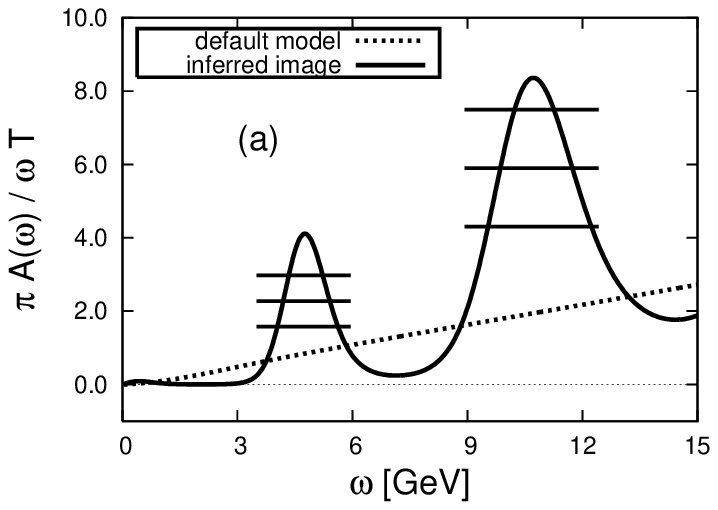}
      \end{minipage}
      \begin{minipage}{1.0\linewidth}
        \includegraphics[width=\linewidth]{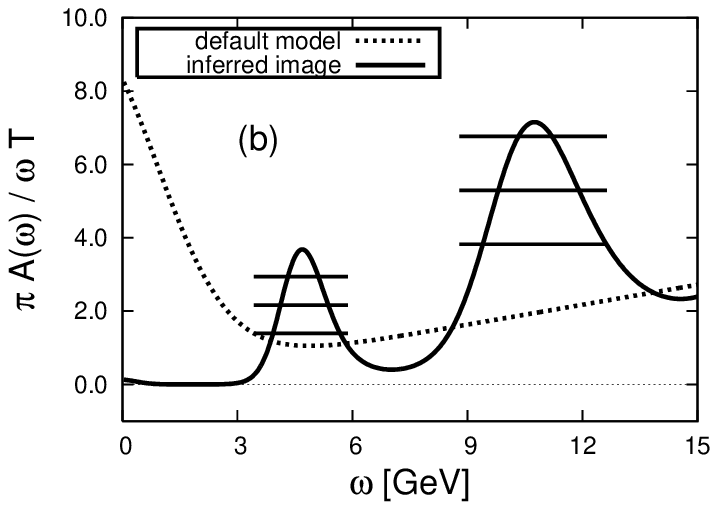}
      \end{minipage}
    \end{center}
    \caption{
      (a) The real part of the complex admittance
      $\pi \, A(\omega) / \omega T$ inferred using the improved MEM analysis
      based on $m(\omega) = m_\mathrm{mono}(\omega)$ in
      Eq.(\ref{eq:3-006}).
      (b) The result of the improved MEM analysis
      based on $m(\omega) = m_\mathrm{lore}(\omega)$ in Eq.(\ref{eq:007})
      with
      $\omega_0 = 1.0 \, \mathrm{GeV}$,
      $\Delta\omega = 2.0 \, \mathrm{GeV}$,
      $\chi = 5.0 \, \mathrm{GeV}$ and
      $\tau_J = 0.1 \, \mathrm{GeV}^{-1}$.
      The dash line denotes the default model $m(\omega) / T$
      and
      the solid line the inferred
      $\pi \, A(\omega) / \omega T$.
      The horizontal position and length of the bars
      indicate the region over which
      $\pi A(\omega) / \omega T$ is averaged,
      while the vertical height of the bars
      denotes the standard deviation in the averaged value
      of $\pi A(\omega) / \omega T$ in the interval.
    }
    \label{fig:001}
  \end{figure}

  The improved MEM analysis is done
  using
  the lattice data $D(\tau)$ ($\tau = \tau_i, \cdots, \tau_f$)
  with $\tau_i/a_\tau = 4$ and $\tau_f/a_\tau = 42$
  and
  the default models $m_\mathrm{mono}(\omega)$ in Eq.(\ref{eq:3-006}) and
  $m_\mathrm{lore}(\omega)$ in Eq.(\ref{eq:007}).
  In Fig.\ref{fig:001}, we show the inferred $\pi \, A(\omega) / \omega$'s,
  which have been checked to be stable
  against the change of $(\tau_i, \tau_f)$ as $(5, 41)$, $(6, 40)$ and $(7, 39)$.
  Figure \ref{fig:001}-(a) indicates that
  $\pi \, A(\omega) / \omega$ has
  no strength in the low-frequency region
  $\omega < 3 \, \mathrm{GeV}$
  and the peak structure with large width around $\omega \sim 4.5 \, \mathrm{GeV}$.
  Notice that due to $m_\mathrm{mono}(\omega)$ having no strength
  in $\omega < 3 \mathrm{GeV}$ and taking a zero value at $\omega = 0$,
  we cannot get any information about
  the applicability of the Israel-Stewart equation to the
  baryon-number dissipative process of the charm quark-gluon plasma
  and the smallness of
  the transport coefficient
  of the baryon-number dissipative process $\chi$ from this result.
  On the other hand,
  the peak structure with large width around $\omega \sim 4.5 \, \mathrm{GeV}$
  can be considered to be a signal of the charmed vector hadron $J/\Psi$
  surviving above $T_C$.
  This result is in good agreement with those of the several groups
  \cite{mem003,mem004,mem005,mem006}.
  Figure \ref{fig:001}-(b) shows that
  $\pi \, A(\omega) / \omega$ is quite different
  from $m_\mathrm{lore}(\omega)$,
  the Lorentzian function in the low-frequency region.
  This result suggests that
  the applicability of the Israel-Stewart equation
  to the baryon-number dissipative process of the charm quark-gluon plasma
  is questionable.
  We have checked the stability
  of the inferred $\pi \, A(\omega) / \omega$
  against the change of
  $\omega_0$,
  $\Delta\omega$,
  $\chi$ and
  $\tau_J$.

\begin{figure}[tHb]
    \begin{center}
      \begin{minipage}{1.0\linewidth}
        \includegraphics[width=\linewidth]{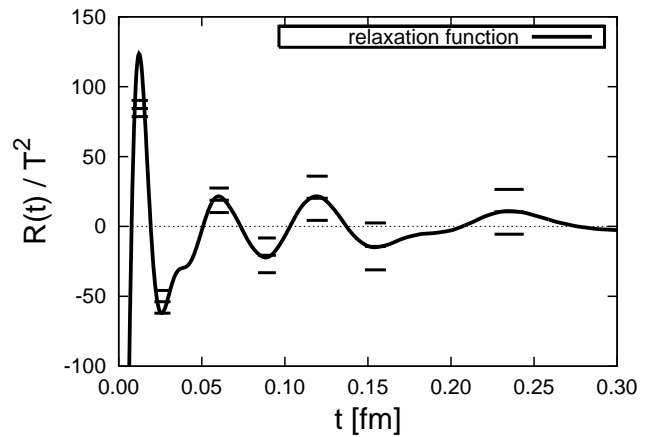}
      \end{minipage}
    \end{center}
    \caption{
      The solid line denotes the relaxation function $R(t) / T^2$
      with $T = 1.62 \, T_C$.
      The meaning of the horizontal position and length of the bars
      is explained in the caption of Fig.\ref{fig:001}.
      $R(t) / T^2$ exhibits a strongly-oscillation damping behaviour
      within the error bars,
      instead of
      the exponentially damping behaviour
      derived from the Israel-Stewart equation.
    }
    \label{fig:002}
  \end{figure}

  We calculate the relaxation function
  expected to be quite different from the exponentially damping function.
  For this purpose,
  we carry out the Fourier transformation of
  the inferred $\pi \, A(\omega) / \omega$ in Fig.\ref{fig:001}-(a).
  Since $\pi \, A(\omega) / \omega$ exhibits
  a linear divergence of $\omega$
  for $\omega\rightarrow \infty$,
  the numerical Fourier transformation is unstable.
  To avoid this numerical instability,
  we convert
  Eq.(\ref{eq:3-003})
  into
  \begin{eqnarray}
    \label{eq:008}
    R(t) =  2 \, \theta(t) \, \Bigg\{ - \frac{m_0}{t^2}
    \nonumber\\
    {} + \frac{1}{\pi} \, \int_0^{\omega_\mathrm{cut}} \!\!
    \mathrm{d}\omega \,
    \cos\omega t \,
    \Bigg[ \frac{\pi}{\omega} \,
      A(\omega) - m_\mathrm{mono}(\omega) \Bigg] \Bigg\},
  \end{eqnarray}
  where we have used $A(\omega) = -A(-\omega)$ and
  the following relation,
  \begin{eqnarray}
    \label{eq:3-007}
    \int_0^\infty\!\!\mathrm{d}\omega \,
    \mathrm{e}^{-i\omega t} \, \omega = - \frac{1}{t^2} \,\,\,\mathrm{for}\,\,\, t \neq 0.
  \end{eqnarray}
  In Eq.(\ref{eq:008}),
  $\omega_\mathrm{cut}$ denotes the cut-off frequency defined as
  $\pi \, A(\omega) / \omega - m_\mathrm{mono}(\omega) \sim 0$
  for $\omega > \omega_\mathrm{cut}$.
  Here we use $\omega_\mathrm{cut} = 100 \, \mathrm{GeV} > \pi /
  a_\tau = 63  \, \mathrm{GeV}$.
  In Fig.\ref{fig:002}, we show the calculated $R(t)$
  together with the error bars.
  The singularity of $R(t=0)$ is inevitable
  because it is originated from
  the divergence of
  $\pi \, A(\omega) / \omega$ at $\omega \rightarrow \infty$
  due to
  the composite operator
  $J^\mu(-i\tau,\,\Vec{x}) = \sum_{f=1}^{N_\mathrm{F}} \,
  \sum_{c=1}^{N_\mathrm{C}} \, \bar{\psi}_{fc}(-i\tau,\,\Vec{x}) \, \gamma^\mu \,
  \psi(-i\tau,\,\Vec{x})_{fc}$
  being ill-defined at the same coordinate.
  Therefore, we concentrate on $R(t)$
  for $t > a_\tau = 9.75 \times 10^{-3} \, \mathrm{fm}$.
  As expected,
  $R(t)$ is not a exponential damping function
  but a strongly-oscillation damping function.
  This oscillation damping behaviour of $R(t)$ is attributed to
  the peak structure with the finite width in $\pi \, A(\omega) / \omega$
  correspondent to the surviving $J/\Psi$ at $T = 1.62 \, T_C$.
  The relaxation time $\tau_\mathrm{osc}$ of the oscillation damping function
  is approximated by the inverse of the width of the low-lying peak structure.
  Accordingly,
  we can estimate
  $\tau_\mathrm{osc} \sim 0.1 \, \mathrm{fm}$
  from Fig.\ref{fig:001}-(a),
  which seems to be consistent with $R(t)$ in FIG. \ref{fig:002}.
  The disappearance of $J/\Psi$ at the higher temperature
  would drive the drastic change of $R(t)$ and $\tau_\mathrm{osc}$.
  
  \subsection{Relaxation Equation Applicable to Strongly Interacting System\label{sec:003C}}
  Finally we discuss the form of a relaxation equation
  corresponding to the spectral function
  with the peak structures in the finite-frequency region.
  Here we consider the spectral function
  for the shear viscous dissipative process
  at the rest frame as a typical example,
  \begin{eqnarray}
    \label{eq:3-008}
    \frac{\pi}{\omega} \, A(\omega,\,\Vec{k}) =
    \frac{\eta\,(\omega^2_\pi + \tau_\pi^{-2})/2}{(\omega - \omega_\pi)^2 + \tau_\pi^{-2}}
  + \frac{\eta\,(\omega^2_\pi + \tau_\pi^{-2})/2}{(\omega + \omega_\pi)^2 + \tau_\pi^{-2}},
  \end{eqnarray}
  where
  $\omega_\pi(\ge 0)$ denotes the peak position.
  Notice that Eq.(\ref{eq:3-008})
  agrees with Eq.(\ref{eq:2-005}) by the setting of $\omega_\pi = 0$.
  Using Eqs.(\ref{eq:admittance}) and (\ref{eq:3-008}),
  we have the correspondent complex admittance,
  \begin{eqnarray}
    \label{eq:3-009}
    \tilde{R}(k^0,\,\Vec{k}) = \eta \, \frac{\omega^2_\pi +
      \tau_\pi^{-2}}{\tau_\pi^{-1}} \, \frac{- i k^0 + \tau_\pi^{-1}}{\omega^2_\pi + (-
      i k^0 + \tau_\pi^{-1})^2}.
  \end{eqnarray}
  By combining Eqs.(\ref{eq:2-004}) and (\ref{eq:3-009}),
  we find the form of the relaxation equation
  corresponding to the spectral function
  with the two peak structures at $\omega = \pm \omega_\pi$
  to be
  \begin{eqnarray}
    \label{eq:3-010}
    \Big[
      \omega^2_\pi \, \tau^2_\pi + (1 + \tau_\pi \, \partial_t)^2
      \Big] \, J(t,\,\Vec{x})\nonumber\\
    = \eta \, (\omega^2_\pi \, \tau^2_\pi + 1)
    \, (1 + \tau_\pi \, \partial_t) \, X(t,\,\Vec{x}).
  \end{eqnarray}
  It is found that
  the temporal derivative $\partial_t$ appears
  also in the right hand side of Eq.(\ref{eq:3-010}),
  in contrast with
  the Israel-Stewart-type relaxation equation (\ref{eq:2-008}).
  We identify Eq.(\ref{eq:3-010})
  as a more probable form of the relaxation equation
  applicable to the strongly interacting system
  including QGP slightly above $T_C$,
  of which spectral function has peak structures
  in the finite-frequency region.
  
  As shown in subsection \ref{sec:003B},
  the baryon-number dissipative process of the charm quark-gluon plasma
  cannot be described by the Israel-Steart equation.
  We should carry out the causal hydrodynamic simulations
  by incorporating
  the relaxation equation based on
  Eq.(\ref{eq:3-010}) not Eq.(\ref{eq:2-008})
  into the continuity equation (\ref{eq:3-001}).
  
  \section{Summary and Concluding Remarks\label{sec:004}}
  First, in this paper,
  we have pointed out that
  the Israel-Stewart causal hydrodynamic equation is acceptable
  only as long as the relaxation function is a exponentially damping function,
  and the relaxation time
  can be obtained as its damping time constant.
  Using the linear response theory,
  we have derived the formula
  where the relaxation function is represented by the spectral function
  for the conserved current,
  which can be calculated from the microscopic dynamics.
  Furthermore, we have discussed the systematic way
  to construct a causal hydrodynamic equation
  corresponding to the calculated relaxation function.
  
  Then, taking the baryon-number dissipative process
  of the charm quark-gluon plasma for a simple example,
  we have calculated the relaxation function from the spectral function obtained
  using the quenched lattice QCD together with the maximum entropy method.
  The obtained relaxation function 
  shows the strongly-oscillation damping behaviour within the error bars,
  not the exponential damping.
  The origin of the strongly-oscillation damping behaviour
  is the charmed vector hadron $J/\Psi$ surviving at $T = 1.62 \, T_C$.
  This fact suggests that
  the applicability of the Israel-Stewart equation
  to the baryon-number dissipative process of the charm quark-gluon plasma
  is questionable.
  We have shown an idea for the improvement of the Israel-Stewart equation
  by deriving the relaxation equation
  consistent with  the strongly-oscillation damping relaxation function.

  We conclude that
  the Israel-Stewart equation is not unique
  as the causal and dissipative hydrodynamic equation.
  By calculating the relaxation function from the spectral function
  with use of the formula in Eq.(\ref{eq:005}),
  we can investigate the applicability of the Israel-Stewart equation.
  It is noteworthy that
  the form of the causal hydrodynamic equation with dissipative
  effects applicable to QGP
  is determined by the shape of the spectral function
  reflecting the hadronic excitations in QGP.
  This indicates
  an important link between
  the non-equilibrium properties of QGP
  and the hadron spectroscopy at the finite temperature.
  
  In the next paper \cite{nex002},
  we will investigate
  the properties of the relaxation equation
  corresponding to the spectral function
  with the peak structures in the finite-frequency region.
  It seems an interesting subject to
  derive
  the relativistic hydrodynamic equation
  by combining
  the relaxation equation shown in Eq.(\ref{eq:3-010})
  and the continuity equation
  based on the energy-flow equation by Landau \cite{1st002}
  or the particle-flow equation by Tsumura \textit{et al.} \cite{key001,key002},
  which have a stable equilibrium solution.
  
  To discuss the more realistic system than the charm quark-gluon plasma,
  we must investigate the contribution of up, down
  and strange quarks
  to the spectral function for the baryon-number conserved current.
  The light vector hadrons composed of these quarks slightly above $T_C$
  seem to play a more essential role for the slow and long-wavelength fluid behaviour of QGP
  than $J/\Psi$.
  The answer to this conjecture will be reported elsewhere.
  In addition,
  the calculation of
  the relaxation function and/or the spectral function for
  the shear and bulk viscous dissipative process of QGP
  is an interesting future study
  using the lattice QCD with the maximum entropy method
  or another non-perturbative method,
  such as the QCD sum rule or the AdS/CFT duality.
  
  \begin{acknowledgements}
    The author is grateful to Prof. T. Kunihiro,
    Prof. H. Suganuma
    and Dr. T. Kojo for their useful comments and discussions.
    The lattice QCD calculations have been performed on
    NEC-SX5 at Osaka University.
  \end{acknowledgements}
  
  \appendix
  
  \section{Microscopic Representation of Relaxation Function\label{sec:005}}
  In this section,
  we show the derivation of Eq.(\ref{eq:003})
  based on the linear response theory \cite{lrt001}
  together with the non-equilibrium statistical operator method proposed by
  Zubarev \cite{lrt002}.
  
  We consider a statistical operator
  $\hat{\rho}(t) \equiv |\,\Psi(t)\,\rangle \langle\,\Psi(t)\,|$,
  where $|\,\Psi(t)\,\rangle$ denotes a state of the system.
  The time evolution of $\hat{\rho}(t)$ is governed
  by the Liouville equation,
  \begin{eqnarray}
    \label{eq:app01}
    \partial_t \hat{\rho}(t) + i \, [ \hat{H}\,,\,
      \hat{\rho}(t) ] = 0,
  \end{eqnarray}
  or the Neumann equation, i.e., the equation converted from Eq.(\ref{eq:app01}),
  \begin{eqnarray}
    \label{eq:app02}
    \partial_t \ln \hat{\rho}(t) + i \, [ \hat{H}\,,\,\ln
      \hat{\rho}(t) ] = 0,
  \end{eqnarray}
  both of which are equivalent to
  the Schr\"{o}dinger equation,
  $i \, \partial_t |\Psi(t)\rangle = \hat{H} \, |\Psi(t)\rangle$
  with $\hat{H}$ being the Hamiltonian.
  It is noted that
  Eqs.(\ref{eq:app01}) and (\ref{eq:app02})
  hold the time-reversal symmetry.
  Therefore,
  the entropy of the system defined by
  $S(t) \equiv  \mathrm{Tr} \hat{\rho}(t) \, \ln \hat{\rho}(t)$
  is normally conserved.
  
  To obtain the non-equilibrium statistical operator
  $\hat{\rho}_\mathrm{neq}(t)$,
  which exhibits the increase of the entropy,
  the breaking
  of the time-reversal symmetry
  is needed.
  In Ref.\cite{lrt002}, Zubarev has proposed
  to construct
  $\hat{\rho}_\mathrm{neq}(t)$
  as the solution of
  the Neumann equation (\ref{eq:app02})
  under the time-reversal asymmetric boundary condition
  $\hat{\rho}(t\rightarrow -\infty) =
  \hat{\rho}_\mathrm{qeq}(t\rightarrow -\infty|0)$,
  where $\hat{\rho}_\mathrm{qeq}(t|0)$
  denotes the quasi-equilibrium statistical operator
  defined by
  \begin{eqnarray}
    \label{eq:app03}
    \hat{\rho}_\mathrm{qeq}(t_1|t_2) \propto
    \exp\Bigg[-\int\!\!\mathrm{d}^3\Vec{x} \, a_\mu(t_1,\,\Vec{x}) \,
    \hat{T}^{\mu 0}(t_2,\,\Vec{x})\Bigg],
  \end{eqnarray}
  with
  $\mathrm{Tr} \, \hat{\rho}_\mathrm{qeq}(t_1|t_2) = 1$.
  Here, $\hat{T}^{\mu\nu}(t,\,\Vec{x})$ denotes
  a energy-momentum tensor operator
  in the Heisenberg picture,
  which agrees with that of the Schrodinder picture at $t=0$.
  $a_\mu(t,\,\Vec{x})$
  denotes a Lagrange multiplier
  characterizing the system.
  The physical meaning of $a_\mu(t,\,\Vec{x})$
  becomes clear
  from
  the setting of
  $a_\mu(t,\,\Vec{x}) = g_{\mu 0} / T$,
  which
  reproduces
  the equilibrium statistical operator
  $\hat{\rho}_\mathrm{eq}
  \equiv \mathrm{e}^{-\hat{H}/T} \, / \, \mathrm{Tr} \,
  \mathrm{e}^{-\hat{H}/T}$
  with
  $\hat{H} = \int\!\!\mathrm{d}^3\Vec{x}\,\hat{T}^{00}(0,\,\Vec{x})$.
  
  It is noted that
  imposing
  the time-reversal asymmetric boundary condition
  $\hat{\rho}(t\rightarrow -\infty) =
  \hat{\rho}_\mathrm{qeq}(t\rightarrow -\infty|0)$
  to Eq.(\ref{eq:app02})
  is equivalent to
  adding a time-reversal asymmetric source term
  to the right hand side of Eq.(\ref{eq:app02}),
  \begin{eqnarray}
    \label{eq:app04}
    \partial_t \ln \hat{\rho}(t) + i \, [ \hat{H}\,,\,\ln
      \hat{\rho}(t) ] = - \epsilon \, \ln \Big[ \hat{\rho}(t)
      \Big/ \hat{\rho}_\mathrm{qeq}(t|0) \Big],
  \end{eqnarray}
  where
  $\epsilon$ denotes an infinitesimal positive constant
  and
  the limit of $\epsilon \rightarrow +0$ follows
  the thermodynamical limit.
  $\hat{\rho}_\mathrm{neq}(t)$
  can be obtained as
  $\hat{\rho}_\mathrm{neq}(t)
  = \lim_{\epsilon\rightarrow +0} \, \hat{\rho}_\epsilon(t)$,
  where
  $\hat{\rho}_\epsilon(t)$
  denotes
  the solution of Eq.(\ref{eq:app04})
  and $\lim_{\epsilon \rightarrow +0}$
  is suppressed later on.
  The solution of Eq.(\ref{eq:app04}) reads
  \begin{eqnarray}
    \label{eq:app05}
    \ln \hat{\rho}_\mathrm{neq}(t) = 
      \epsilon \, \int_{-\infty}^t\!\!\mathrm{d}s \,
      \mathrm{e}^{\epsilon (s-t)} \, \ln
      \hat{\rho}_\mathrm{qeq}(s|s-t).
  \end{eqnarray}
  A partial integration of
  the right hand side of Eq.(\ref{eq:app05})
  leads us to
  \begin{eqnarray}
    \label{eq:app24}
    \hat{\rho}_\mathrm{neq}(t) =
    \mathrm{e}^{\hat{A}+\hat{B}} \,/\,
    \mathrm{Tr}\,\mathrm{e}^{\hat{A}+\hat{B}},
  \end{eqnarray}
  where
  \begin{eqnarray}
    \label{eq:app06}
    \hat{A} \equiv -\int\!\!\mathrm{d}^3\Vec{x} \, a_\mu(t,\,\Vec{x}) \,
    \hat{T}^{\mu 0}(0,\,\Vec{x}), \\
    \label{eq:app07}
    \hat{B} \equiv \int_{-\infty}^t\!\!\mathrm{d}s \,
    \mathrm{e}^{\epsilon (s-t)} \!\!
    \int\!\!\mathrm{d}^3\Vec{x} \, \partial_\mu a_\nu(s,\,\Vec{x})
    \, \hat{T}^{\mu\nu}(s-t,\,\Vec{x}).
  \end{eqnarray}
  Here we have used $\partial_\mu \hat{T}^{\mu\nu} = 0$.
  
  Notice that
  $\hat{B}$ contains
  $\partial_\mu a_\nu$,
  i.e., the gradient of the Lagrange multiplier characterizing the system.
  We shall consider
  the system not so far from
  the equilibrium state,
  where
  $\partial_\mu a_\nu$ is small.
  It is important that
  under this consideration
  we can regard $\hat{B}$
  as a perturbation term to $\hat{A}$
  and utilize the linear response theory,
  where
  $\partial_\mu a_\nu$ is
  identified as an external field.
  We expand $\hat{\rho}_\mathrm{neq}(t)$
  up to the first order of $\hat{B}$ as follows,
  \begin{eqnarray}
    \label{eq:app08}
    \hat{\rho}_\mathrm{neq}(t) =
    \Big( 1 + \int_0^1\!\!\mathrm{d}\lambda \,
    \mathrm{e}^{+\lambda\hat{A}} \, (\hat{B} - {\langle \, \hat{B} \,
      \rangle}_A) \, \mathrm{e}^{-\lambda\hat{A}}\Big) \, \hat{\rho}_A,
  \end{eqnarray}
  where
  $\hat{\rho}_A \equiv \mathrm{e}^{\hat{A}} \,/\, \mathrm{Tr}\,\mathrm{e}^{\hat{A}}$
  and
  ${\langle \, \hat{O} \, \rangle}_A \equiv \mathrm{Tr}\,\hat{\rho}_A
  \, \hat{O}$
  with an arbitrary operator $\hat{O}$.
  Notice that
  $\hat{\rho}_A = \hat{\rho}_\mathrm{qeq}(t|0)$.
  
  Using Eq.(\ref{eq:app08}),
  we calculate the energy-momentum tensor realized in the
  non-equilibrium state,
  \begin{eqnarray}
    \label{eq:app09}
    T^{\mu\nu}(t,\,\Vec{x}) \equiv \mathrm{Tr} \,
    \hat{\rho}_\mathrm{neq}(t) \,
    \hat{T}^{\mu\nu}(0,\,\Vec{x}).
  \end{eqnarray}
  We decompose
  $T^{\mu\nu}(t,\,\Vec{x})$
  into
  the unperturbative term and the perturbative one
  as
  $T^{\mu\nu}(t,\,\Vec{x}) = T^{\mu\nu}_\mathrm{0}(t,\,\Vec{x})
  + \delta T^{\mu\nu}(t,\,\Vec{x})$,
  where
  \begin{eqnarray}
    \label{eq:app10}
    T^{\mu\nu}_\mathrm{0}(t,\,\Vec{x}) &\equiv&
    {\langle \, \hat{T}^{\mu\nu}(0,\,\Vec{x}) \, \rangle}_A,\\
    \label{eq:app11}
    \delta T^{\mu\nu}(t,\,\Vec{x})
    &\equiv& \int\!\!\mathrm{d}u\,\mathrm{d}^3\Vec{y} \,
    R^{\mu\nu\rho\sigma}(t,\,\Vec{x}|u,\,\Vec{y})\nonumber\\
    &&{}\times\partial_\rho a_\sigma(u,\,\Vec{y}).
  \end{eqnarray}
  Here $T^{\mu\nu}_{\mathrm{0}}$ represents
  the perfect-fluid part of
  the energy-momentum tensor,
  while $\delta T^{\mu\nu}$
  the dissipative part.
  $R^{\mu\nu\rho\sigma}(t,\,\Vec{x}|u,\,\Vec{y})$ in Eq.(\ref{eq:app11})
  is a relaxation function defined by
  \begin{eqnarray}
    \label{eq:app12}
    &&R^{\mu\nu\rho\sigma}(t,\,\Vec{x}|u,\,\Vec{y}) =
    \theta(t-u) \, \mathrm{e}^{-\epsilon(t-u)}\nonumber\\
    &&{}\times\Bigg[
      \int_0^1\!\!\mathrm{d}\lambda \, 
	  {\langle \, 
	    \hat{T}^{\mu\nu}(0,\,\Vec{x}) \,
	    \mathrm{e}^{+\lambda\hat{A}} \,
	    \hat{T}^{\rho\sigma}(u - t,\,\Vec{y}) \, \mathrm{e}^{-\lambda\hat{A}}
	    \, \rangle}_A \nonumber\\
	  &&-{\langle \, 
	    \hat{T}^{\mu\nu}(0,\,\Vec{x}) \, \rangle}_A
	  \, {\langle \,
	    \hat{T}^{\rho\sigma}(u - t,\,\Vec{y})
	    \, \rangle}_A
	  \Bigg].
  \end{eqnarray}
  Equation (\ref{eq:app12}) is one of the main results in this section.
  
  As the thermal average used in Eq.(\ref{eq:app12}) is
  defined with use of $\hat{\rho}_A = \hat{\rho}_\mathrm{qeq}(t|0)$
  not $\hat{\rho}_\mathrm{eq}$,
  Eq.(\ref{eq:app12}) seems inconvenient
  to calculate
  $R^{\mu\nu\rho\sigma}(t,\,\Vec{x}|u,\,\Vec{y})$
  employing a technique established in the finite-temperature field theory.
  To derive
  $R^{\mu\nu\rho\sigma}(t,\,\Vec{x}|u,\,\Vec{y})$ written by
  $\hat{\rho}_\mathrm{eq}$,
  we shall make an approximation for Eq.(\ref{eq:app11}).
  We consider a situation
  where a fluctuation of $a_\mu(t,\,\Vec{x})$
  is small.
  Notice that this approximation
  differs from that used
  in the linear response theory,
  and is implemented
  in a systematic way
  using the expansion,
  \begin{eqnarray}
    \label{eq:app13}
    a_\mu(t,\,\Vec{x}) = \bar{a}_\mu + \varepsilon \, \delta a_\mu(t,\,\Vec{x}),
  \end{eqnarray}
  where $\varepsilon$ is an infinitesimal constant
  different from $\epsilon$.
  We evaluate a leading order of $\delta T^{\mu\nu}(t,\,\Vec{x})$
  with respect to $\varepsilon$.
  Due to $\partial_\rho a_\sigma(u,\,\Vec{y})$
  being the first order of $\varepsilon$,
  the zeroth-order evaluation of $R^{\mu\nu\rho\sigma}(t,\,\Vec{x}|u,\,\Vec{y})$
  is sufficient.
  
  First,
  using $a_\mu(t,\,\Vec{x}) = \bar{a}_\mu$,
  we have
  \begin{eqnarray}
    \label{eq:app14}
    \hat{A} = - \bar{a}_\mu \, \hat{P}^\mu,
  \end{eqnarray}
  where the energy-momentum vector operator
  $\hat{P}^\mu \equiv \int\!\!\mathrm{d}^3\Vec{x} \, \hat{T}^{\mu
  0}(0,\,\Vec{x})$.
  With use of
  the unitary transformation for
  the temporal and spatial translation
  by $\hat{P}^\mu$,
  \begin{eqnarray}
    \label{eq:app15}
    \mathrm{e}^{+\lambda\hat{A}} \,
    \hat{O}(t,\,\Vec{x}) \,
    \mathrm{e}^{-\lambda\hat{A}}
    = \hat{O}(t + i \, \lambda \,
    \bar{a}^0,\,\Vec{x} + i \, \lambda \, \Vec{\bar{a}}),
  \end{eqnarray}
  and
  the cluster property of the correlation function,
  \begin{eqnarray}
    \label{eq:app16}
    {\langle \, 
      \hat{O}(0,\,\Vec{x}) \, \hat{O}(t,\,\Vec{y})
      \, \rangle}_A
    = {\langle \, 
    \hat{O}(0,\,\Vec{x})
    \, \rangle}_A
    \, {\langle \, \hat{O}(t,\,\Vec{y})
      \, \rangle}_A\nonumber\\
    \,\,\,\mathrm{for}\,\,\,
    |t|\rightarrow\infty,
  \end{eqnarray}
%  we can convert the square bracket in Eq.(\ref{eq:app12}) into
  we can convert the right hand side of Eq.(\ref{eq:app12}) into
  \begin{eqnarray}
    \label{eq:app17}
    \theta(t-u) \, \mathrm{e}^{-\epsilon(t-u)} \,
    \int_0^1\!\!\mathrm{d}\lambda \, 
    \int_{-\infty}^0\!\!\mathrm{d}s \, 
    \langle \, \hat{T}^{\mu\nu}(0,\,\Vec{x})\nonumber\\
    \times \, {\partial_s \hat{T}^{\rho\sigma}(u - t + (i \, \lambda
      + s) \,
      \bar{a}^0,\,\Vec{y} + (i \, \lambda + s) \, \Vec{\bar{a}})
      \, \rangle}_A.
  \end{eqnarray}
  By replacing
  $\partial_s \rightarrow -i \, \partial_\lambda$,
  integrating with respect to $\lambda$
  and using the translational invariance,
  we find Eq.(\ref{eq:app17}) to be
  \begin{eqnarray}
    \label{eq:app18}
    \theta(t-u) \, \mathrm{e}^{-\epsilon(t-u)} \,
    \int_{-\infty}^0\!\!\mathrm{d}s\nonumber\\
    {}\times i \,
    {\langle \, 
      [ \hat{T}^{\mu\nu}(t-u,\,\Vec{x}-\Vec{y}) \,,\,
	\hat{T}^{\rho\sigma}(s \, \bar{a}^0,
	\,s \, \Vec{\bar{a}}) ]
      \, \rangle}_A.
  \end{eqnarray}
  Using Eq.(\ref{eq:app18}),
  we can rewrite Eq.(\ref{eq:app11}) as
  a Lorentz-covariant form,
  \begin{eqnarray}
    \label{eq:app19}
    \delta T^{\mu\nu}(x) 
    &=& \int\!\!\mathrm{d}^4y \,
    R^{\mu\nu\rho\sigma}(x - y)\,
    \partial_\rho a_\sigma(y),\\
    \label{eq:app20}
    R^{\mu\nu\rho\sigma}(x)
    &\equiv& \theta(x^0) \, \mathrm{e}^{-\epsilon x^0}
    \, \int_{-\infty}^0\!\!\mathrm{d}s\nonumber\\
    & &{}\times i \, {\langle \, 
      [ \hat{T}^{\mu\nu}(x) \,,\,
	\hat{T}^{\rho\sigma}(s \, a) ]
      \, \rangle}_A.
  \end{eqnarray}
  
  Then,
  using $\bar{a}_\mu = u_\mu \,/\,T$ ($u_\mu \, u^\mu = 1$),
  we derive $R^{\mu\nu\rho\sigma}(x)$
  written by $\hat{\rho}_\mathrm{eq}$.
  For this purpose,
  we utilize a decomposition
  of $\delta T^{\mu\nu}(x)$
  based on the irreducible tensors $u^\mu$,
  $\Delta^{\mu\nu}=g^{\mu\nu} - u^\mu \, u^\nu$
  and
  $\Delta^{\mu\nu\rho\sigma} = 1/2 \, (\Delta^{\mu\rho}
  \,\Delta^{\nu\sigma} + \Delta^{\mu\sigma} \,\Delta^{\nu\rho} - 2/3 \,
  \Delta^{\mu\nu} \,\Delta^{\rho\sigma})$,
  namely,
  \begin{eqnarray}
    \delta T^{\mu\nu}(x) = E(x) \, u^\mu \, u^\nu - \Pi(x) \,
    \Delta^{\mu\nu}\nonumber\\
    + Q^\mu(x) \, u^\nu + Q^\nu(x) \, u^\mu + \pi^{\mu\nu}(x),
  \end{eqnarray}
  where
  $E(x) \equiv \delta T^{ab}(x) \, u_a \, u_b$,
  $\Pi(x) \equiv - \delta T^{ab}(x) \, \Delta_{ab} \, / \, 3$,
  $Q_\mu(x) \equiv \delta T^{ab}(x) \, u_a \, \Delta_{b\mu}$
  and
  $\pi_{\mu\nu}(x) \equiv \delta T^{ab}(x) \, \Delta_{ab\mu\nu}$.
  Here we focus the shear viscous stress
  $\pi_{\mu\nu}(x)$.
  Using Curie's principle \cite{mic001} and the spatial isotropy,
  we have
  \begin{eqnarray}
    \label{eq:app20-1}
	  {\langle \, 
	    [ \hat{\pi}^{\mu\nu}(x) \,,\,
	      \hat{T}^{\rho\sigma}(s \, u / T) ]
	    \, \rangle}_A\nonumber\\
	  = \Delta^{\mu\nu\rho\sigma} \, \frac{1}{10} \, {\langle \, 
	    [ \hat{\pi}^{ab}(x) \,,\,
	      \hat{\pi}_{ab}(s \, u / T) ]
	    \, \rangle}_A,
  \end{eqnarray}
  where $\hat{\pi}^{\mu\nu} \equiv \Delta^{\mu\nu a b} \, \hat{T}_{ab}$.
  This relation leads us to
  \begin{eqnarray}
    \label{eq:app21}
    \pi_{\mu\nu}(x)
    &=& \Delta_{\mu\nu\rho\sigma} \, \int\!\!\mathrm{d}^4y \,
    R(x - y)\,
    2 \, \partial^\rho u^\sigma(y),\\
    \label{eq:app22}
    R(x) &\equiv& \theta(x^0) \, \mathrm{e}^{-\epsilon x^0} \,
    \int_{-\infty}^0\!\!\mathrm{d}s \nonumber\\
    &&{}\times i \, \frac{1}{10 \, T} \, {\langle \, 
      [ \hat{\pi}^{\mu\nu}(x) \,,\,
	\hat{\pi}_{\mu\nu}(s \, u/T) ]
      \, \rangle}_A.
  \end{eqnarray}
  It is noted that,
  due to the integrand of $R(x)$ in Eq.(\ref{eq:app22})
  being a Lorentz-scalar function,
  the integrand takes the same value for an arbitrary $u^\mu$.
  Therefore,
  we can set $u^\mu = (1,\,0,\,0,\,0)$ in Eq.(\ref{eq:app22})
  without loss of generality,
  and using the spatial isotropy we arrive at
  \begin{eqnarray}
    \label{eq:app23}
    R(x) &=& R(t,\,\Vec{x})\nonumber\\
    &=& \theta(t) \, \mathrm{e}^{-\epsilon t}
    \, \int_{-\infty}^0\!\!\mathrm{d}s\nonumber\\
    &&{} \times \theta(t - s) \, i \,
    \langle \, 
      [ \hat{T}_{12}(t,\,\Vec{x}) \,,\,
	\hat{T}_{12}(s,\, \Vec{0}) ]
      \, \rangle,
  \end{eqnarray}
  where $\langle \,\hat{O} \, \rangle \equiv \mathrm{Tr} \,
  \hat{\rho}_\mathrm{eq} \, \hat{O}$.
  Notice that
  the integrand of Eq.(\ref{eq:app23})
  is identical to
  the retarded Green function
  which has the spectral representation as shown in Eq.(\ref{eq:004}).
  Equation (\ref{eq:app23}) is the other of the main results in this section.

\end{document}